\documentclass[useAMS,usenatbib]{mn2e}
\usepackage{graphicx}
 
\def\be{\begin{equation}} 
\def\ee{\end{equation}}

\def\HI{\hbox{H~$\scriptstyle\rm I\ $}} 
\def\HII{\hbox{H~$\scriptstyle\rm II\ $}}

\def\gsim{\lower.5ex\hbox{\gtsima}} 
\def\lsim{\lower.5ex\hbox{\ltsima}} \def\gtsima{$\; \buildrel > \over 
\sim \;$} \def\ltsima{$\; \buildrel < \over \sim \;$} \def\prosima{$\; 
\buildrel \propto \over \sim \;$} \def\gsim{\lower.5ex\hbox{\gtsima}} 
\def\lsim{\lower.5ex\hbox{\ltsima}} 
\def\simgt{\lower.5ex\hbox{\gtsima}} 
\def\simlt{\lower.5ex\hbox{\ltsima}} 
\def\simpr{\lower.5ex\hbox{\prosima}} \def\la{\lsim} \def\ga{\gsim}

\def\gtsima{$\; \buildrel > \over \sim \;$} 
\def\ltsima{$\; \buildrel < \over \sim \;$} 
\def\gsim{\lower.5ex\hbox{\gtsima}} 
\def\lsim{\lower.5ex\hbox{\ltsima}} 
\def\simgt{\lower.5ex\hbox{\gtsima}} 
\def\simlt{\lower.5ex\hbox{\ltsima}} 
\def\simpr{\lower.5ex\hbox{\prosima}} 
\def\la{\lsim} 
\def\ga{\gsim}

\def\E3{{\cal E}_{\rm g}^{III}}

\title[The visibility of LAEs during reionization]{The visibility of Lyman Alpha Emitters during reionization} 
\author[Dayal, Maselli \& Ferrara]{Pratika Dayal$^{1}$\thanks{E-mail: 
dayal@sissa.it (PD)}, Antonella Maselli $^{2}$ \& Andrea Ferrara$^{3}$    \\ 
$^{{1}}$ SISSA/International School for Advanced Studies, Via Beirut 2-4 Trieste, Italy, 34014\\ 
$^{2}$ Osservatorio Astrofisico di Arcetri, Largo Enrico Fermi 5, 50125 Firenze, Italy \\
$^{3}$ Scuola Normale Superiore, Piazza dei Cavalieri 7, 56126 Pisa, Italy}

\begin{document} 
 
%\date{} %{Received 2010 January 18; in original form 2010 January 18} 
 
\pagerange{\pageref{firstpage}--\pageref{lastpage}} \pubyear{2002} 
 
\maketitle 
 
\label{firstpage} 
\begin{abstract}
We present the first Lyman Alpha Emitter (LAE) study that combines: (i) cosmological SPH
simulations run using {\tt GADGET-2}, (ii) radiative transfer simulations ({\tt CRASH}),
and (iii) a previously developed LAE model. This complete LAE model accounts for the
intrinsic LAE Ly$\alpha$/continuum luminosity, dust enrichment and Ly$\alpha$ transmission
through the intergalactic medium (IGM), to quantify the effects of reionization, dust and
velocity fields on the Ly$\alpha$ and UV Luminosity Functions (LF). We find that a model
neglecting dust sorely fails to reproduce either the slope or the magnitude of the observed
Ly$\alpha$ and UV LFs. Clumped dust is required to simultaneously fit the observed UV and
Ly$\alpha$ LFs, such that the intrinsic Ly$\alpha$-to-continuum luminosity is enhanced by a
factor $f_\alpha/f_c \sim 1.5$ (3.7) excluding (including) peculiar velocities. The higher
value including velocity fields arises since LAEs reside in large potential wells and
inflows decrease their Ly$\alpha$ transmission. For the first time, a {\it degeneracy} is
found between the the ionization state of the IGM and the clumping of dust inside
high-redshift galaxies. The Ly$\alpha$ LF $z \sim 5.7$ can be well reproduced (to within a
$5\sigma$ error) by a wide range of IGM average neutral hydrogen fraction, $3.4 \times
10^{-3} < \langle \chi_{HI} \rangle < 0.16$, provided that the increase in the Ly$\alpha$
transmission through a more ionized IGM  is compensated by a decrease in the Ly$\alpha$ escape fraction
from the galaxy due to dust absorption. The physical properties of LAEs are presented, along with a discussion of the assumptions adopted.

\end{abstract} 


% ************************************************************************** 
\section{Introduction} 
\label{intro}                                                                                                         
The epoch of reionization marks the second major change in the ionization state of the Universe. Reionization begins when the first sources of neutral hydrogen (\HI) ionizing photons form within dark matter potential wells and start building an ionization region around themselves, the so-called Str\"omgren sphere. However, the reionization history and the redshift at which it ends still remain the subject of much discussion. This is because the reionization process depends on a number of parameters including the initial mass function (IMF) of the first sources, their star formation rates (SFR), their stellar metallicity and age, the escape fraction of \HI ionizing photons produced by each source and the clumping of the intergalactic medium (IGM), to name a few.

Given the large number of free parameters that inevitably enter into the construction of theoretical reionization models, it is imperative to compare and update the models as fresh data sets are acquired. In this sense, it has been suggested (Malhotra \& Rhoads 2004, 2005; Santos 2004;  Mesinger, Haiman \& Cen 2004; Haiman \& Cen 2005; Dijkstra, Lidz \& Wyithe 2007;
Mesinger \& Furlanetto 2008; Dayal, Ferrara \& Gallerani 2008; Dayal et al. 2009a, 2009b) that a class of high redshift galaxies called Lyman Alpha Emitters (LAEs) could be an important addition to complementary data sets to constrain the reionization history. 

LAEs, galaxies identified by means of their very strong Ly$\alpha$ line (1216 \AA) emission, have been becoming increasingly popular as probes of reionization for three primary reasons. Firstly, specific signatures like the strength, width and the continuum break bluewards of the Ly$\alpha$ line make the detection of LAEs unambiguous to a large degree. Secondly, since Ly$\alpha$ photons have a large absorption cross-section against \HI, their attenuation can be used to put constraints on the ionization state of the IGM. Thirdly, there are hundreds of confirmed LAEs at $z \sim 4.5$ (Finkelstein et al. 2007), $z \sim 5.7$ (Malhotra et al. 2005; Shimasaku et al. 2006) and $z \sim 6.6$ (Taniguchi et al. 2005; Kashikawa et al. 2006), which exactly probe the redshift range around which reionization is supposed to have ended.

The data accumulated on LAEs shows some very surprising features;
while the apparent Ly$\alpha$ luminosity function (LF) does not show
any evolution between $z=3.1$ - $5.7$ (Ouchi et al. 2008), it changes
appreciably between $z=5.7$ and $6.6$, with $L_*$ (the luminosity of the break, after which the number density decreases rapidly) at $z=6.6$ (Kashikawa et al. 2006) being about 50\% of the value at $z=5.7$ (Shimasaku et al. 2006). Unexpectedly, however, the ultraviolet (UV) LF does not show any evolution between these same redshifts. Although Kashikawa et al. (2006) have proposed this evolution in the Ly$\alpha$ LF to be indicative of a sudden change in the ionization state of the IGM, the problem of why reionization would affect the high luminosity tail of the Ly$\alpha$ LF rather than the faint end, as expected, remains.

In spite of a number of diverse approaches, the effects of the ionization state of the the IGM on the visibility of LAEs still remain poorly understood. This is primarily because understanding/ constraining the ionization state of the IGM using LAEs requires: (a) a detailed knowledge of the physical properties of each galaxy, including the SFR, stellar age and stellar metallicity, needed to calculate the intrinsic Ly$\alpha$ and continuum luminosity produced by stellar sources, (b) the intrinsic Ly$\alpha$ and continuum luminosity produced by the cooling of collisionally excited \HI in the interstellar medium (ISM), (c) an understanding of the dust formation, enrichment and distribution in each galaxy, necessary to calculate the fractions of escaping Ly$\alpha$ and continuum photons, and (d) a full radiative transfer (RT) calculation to obtain the fraction of Ly$\alpha$ luminosity transmitted through the IGM for each galaxy.

Although a number of simulations have been used in the past to study LAEs, they lack one or more of the aforementioned ingredients. Since they used an N-body simulation, McQuinn et al. (2007) were forced to neglect the intrinsic properties of the galaxies and their dust enrichment, although they carried out analytic RT. Iliev et al. (2008) used a simulation that only followed dark matter; they also could not use the intrinsic galaxy properties or calculate the dust enrichment, although they carried out a complete RT calculation. Nagamine et al. (2008) used an SPH (smoothed particle hydrodynamics) simulation to obtain the intrinsic SFR for each galaxy; however, they did not calculate the dust enrichment and assumed a fixed value of the IGM transmission, ignoring RT. Dayal et al. (2009a, 2009b) used an SPH simulation and the intrinsic galaxy properties to calculate the luminosity from both stellar sources and cooling of \HI, calculated the dust enrichment and the IGM transmission; the only missing ingredient in their work was the RT calculation. Most recently, Zheng et al. (2009) have carried out a RT calculation on an SPH simulation; however, they have not included any dust calculation or the luminosity contribution from the cooling of \HI. 

%{\tt TO ADD} most past simulations lack one or the other: McQuinn (analytic RT, no galaxies), Iliev (RT, no galaxies), Nagamine (galaxies, no RT), Cen (RT, galaxies but no dust or gas), we have all (Radiative transfer (RT)). Our aim is to determine how the LF is affected during the progress of (patchy) reionization labeled by different $\langle \chi_{HI} \rangle$, and including clustering of sources. This is the advanced
%version of the argument of Rhoads \& Malhotra 2002, i.e. concentrated on the LF
%rather than on clustering (although we discuss it).

Our aim in this work is to build a LAE model containing all these ingredients, so as to determine the relative importance of dust, peculiar velocity fields and patchy reionization, in shaping the observed Ly$\alpha$ and UV LFs.  We start with an introduction to the SPH and RT simulations used in Sec.~\ref{simulations}. We then present the main ingredients of our LAE model, including the calculation of the intrinsic luminosity from stellar sources/cooling of \HI, the dust enrichment and the IGM transmission, in Sec.~\ref{model}. Once the model is laid down, we present the results obtained with it and quantify the relative importance of dust, peculiar velocities and reionization on the Ly$\alpha$ and UV LFs in Sec.~\ref{reio effects}. The physical properties of the galaxies identified as LAEs are shown in Sec.~\ref{phy prop}. We conclude by mentioning the caveats and shortcomings of our model in Sec.~\ref{conc}. The simulations used in this work are based on a $\Lambda$CDM cosmological model with $\Omega_\Lambda = 0.7$, $\Omega_b = 0.3$, $\Omega_m = 0.04$, $H = 100 h~ {\rm km \, s^{-1} Mpc^{-1}}$, where $h =0.7$ and a scale invariant power spectrum of the initial density perturbations is normalized to $\sigma_8 =0.9$.

% ***************************************************************************
\section{Hydro and radiative transfer}
\label{simulations}

\begin{figure*} 
%  \vspace*{10pt} 
 \center{\includegraphics[scale=0.7]{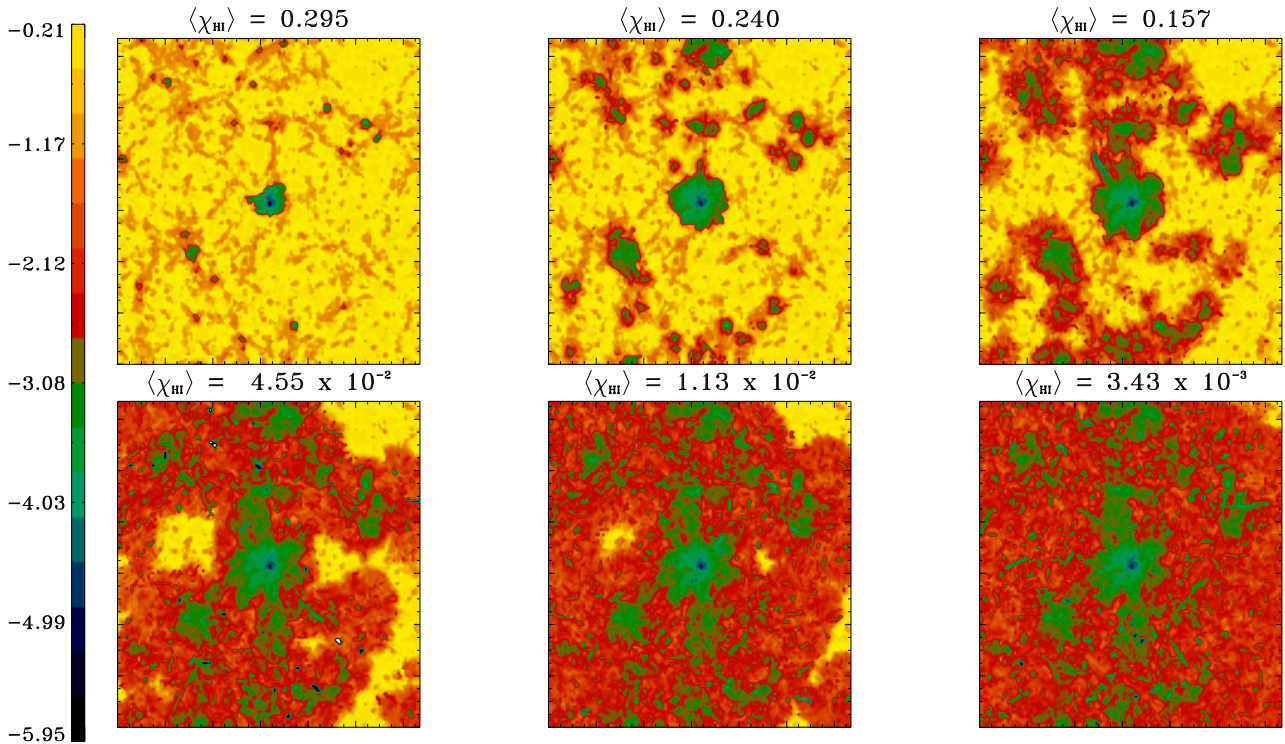}} 
  \caption{Maps (100$h^{-1}$ Mpc on a side) of the spatial distribution of \HI in a 2D cut through the RT simulation box showing the time evolution of the reionization process, with the colorbar showing the values (in log scale) of the \HI plotted. The average decreasing neutral hydrogen fractions marked above each panel, $\langle \chi_{HI} \rangle = 0.295, 0.24, 0.157,4.45\times 10^{-2},1.13 \times 10^{-2} ,3.43 \times 10^{-3}$, correspond to increasing RT simulation timescales of $10, 50, 100, 200, 300, 500$ Myr, respectively.  Details in text of Sec.~\ref{simulations}.}
\label{rt_maps}
\end{figure*}

The Ly$\alpha$ and UV LFs presented in this work are based on the
results from combined runs of SPH and RT simulations, carried out
using {\tt GADGET-2} (Springel 2005)\footnote{http://www.mpa-garching.mpg.de/gadget/}
and {\tt CRASH} (Maselli, Ferrara \& Ciardi 2003; Maselli, Ciardi \&
Kanekar 2009) respectively, which are coupled to a previously developed LAE model (Dayal et al. 2008; Dayal et al. 2009a, 2009b). {\tt GADGET-2} generates the redshift
evolution of the density field, the baryonic density
distribution and the velocity fields in a 100$h^{-1}$ Mpc comoving volume; we obtain a snapshot of the simulation 
at $z \sim 6.1$. The specific simulation used in this work is the G5 run described in
Springel and Hernquist (2003), which is part of an accurate study
focused on modelling the star formation history of the universe. We
have chosen the G5 run since it contains several physical ingredients necessary for our investigation. Firstly, the volume is large enough so that cosmic variance is minimized 
in the determination of the LFs. Secondly, the SFR ($\dot M_*$) is
self-consistently inferred from physically motivated prescriptions
that convert gas particles into stellar particles and that properly take account of the mechanical/chemical feedback associated to supernovae and galactic
winds. This point is
particularly relevant since the intrinsic luminosity of the galaxies, both in the Ly$\alpha$ and UV, depends sensitively on $\dot M_*$. Also, since the physics governing star formation and galaxy evolution is modelled accurately, the simulation gives us a reliable representation of the galaxy population. 
As expected, the large simulation volume naturally leads to a relatively coarse mass resolution with the resolution mass being 
$2.12 \times 10^{10}$ $M_\odot$ ($3.26\times 10^8$ $M_\odot$) for dark
matter (gas particles). Running a friends-of-friends (FOF) algorithm
on the SPH particle distribution, we identify galaxies and obtain
their intrinsic properties, including $\dot M_*$, the mass weighted
stellar metallicity ($Z_*$), the total gas and stellar masses. These
are then used to calculate the intrinsic Ly$\alpha$/continuum
luminosity, the dust enrichment and the escape fraction of
continuum/Ly$\alpha$ photons from the galaxy. All these are presented and discussed in greater detail in Sec.~\ref{model} and \ref{reio effects}.

The RT calculations have been carried out in post-processing mode using the 3D {\tt CRASH} code. The main assumptions made for running {\tt CRASH} are the following: (a) we initially assume, at $z\sim 6.1$,  the gas to be in photoionization equilibrium with a uniform ultraviolet
background (UVB) produced by sources below the simulation resolution scale, corresponding to an average neutral hydrogen fraction, $\langle \chi_{\rm HI} \rangle= 0.3$, (b) we assume that
galaxies have formed stars in the past at an average rate $\dot M_*$, equal to the one derived at $z\sim 6.1$ and (c) since the RT simulation
is carried out on a SPH snapshot which contains about 2500 galaxies,
to reduce the computational running time, we run the RT calculation in
the monochromatic mode, with $h\nu=13.665$ eV. This is a reasonable
assumption since the Ly$\alpha$ LFs depend mostly on the average
ionization fraction and on the 3D topology of the fully ionized
regions; we are therefore not interested in the detailed profiles of the ionization front or a highly accurate estimate of the IGM temperature, to probe which the ionizing radiation of each galaxy would have to be computed from its spectrum. 

The specific photoionization rate is calculated using the expression given in Eq.~12 of Dayal, Ferrara \& Gallerani (2008) and then averaging this value over $Z_*$ and stellar
ages ($t_*$) of the galaxies in the the SPH simulation.  The
contribution from all the galaxies determined under the above-mentioned assumptions is then super-imposed on the average UVB value. Finally, we use $f_{esc}=0.02$ (Gnedin et al. 2008), as a characteristic value
of  the escape fraction of \HI ionizing photons from each galaxy. The total ionizing radiation field, which is the sum of that produced by all galaxies identified in the simulation and the UVB, is then evolved with {\tt CRASH} up to the Hubble time corresponding to $z \sim 6.1$. However, as we are interested in assessing the visibility of LAE as a function of the mean ionization fraction which could correspond to different epochs along any assigned reionization history, no particular meaning should be attached to the redshift parameter. 

In Fig.~\ref{rt_maps} we show maps of the \HI fraction across a 2D cut through the RT simulation box, at different simulation times. 
This time sequence then represents a time-line of the reionization process, due to the galaxy population present within the box. Though an approximation, this description catches many of the important features of the reionization process, including the complex topology produced
by the inhomogeneities in the density field, by the galaxy properties and by their spatial distribution.

Fig.~\ref{rt_maps} clearly shows the reionization process in its main three phases. First, isolated ionized hydrogen (\HII) regions start growing around point sources which are preferentially located along the over-dense filaments of matter. In this stage the size of the isolated \HII regions grows differentially according to the production rate of \HI ionizing
photons from the specific source; the few galaxies with large $\dot M_*$ (e.g. the
largest galaxy located at the center having $\dot M_* \sim 726 \, {\rm M_\odot yr^{-1}}$)
can build an ionized region of about 20 Mpc, in a time as short as 10 Myr (see upper-left panel), which 
grows at a very fast rate due to the vigorous \HI ionizing photon output. If observed at the same evolutionary timescale, due to the smaller \HI ionizing photon budget, the \HII regions of smaller galaxies are confined to sizes of the order of few Mpc ($<4 \, {\rm Mpc}$), which grow slowly. After 10 Myr of continuous star formation activity, the average neutral hydrogen
fraction decreases only to $\langle \chi_{HI} \rangle \sim 0.295$. At this early stage, galaxies with $\dot M_* \leq 25 \, {\rm M_\odot \, yr^{-1}}$, transmit only about 20\% of their Ly$\alpha$ luminosity through the IGM. Consequently the low luminosity end of the LF is depressed with respect to the intrinsic emissivity of these sources. On the other hand, galaxies with larger $\dot M_*$ are able to transmit a larger amount ($\sim 30$\%) of their luminosity through the IGM.

As the \HI ionizing photon production from galaxies continues, the sizes of the \HII regions increase with time. At about $100$ Myr (upper right panel), the isolated \HII regions start overlapping, resulting in an enhancement of the local photo-ionization
rate close to the smaller sources. This allows the ionized regions  
associated to the latter to grow faster, and results in the
build-up of large \HII regions also around smaller galaxies. This results in an enhancement in the transmission of the Ly$\alpha$ luminosity for these sources, such that this value increases to $\sim 40$\%, consequently leading to a boost of the low-luminosity end of the Ly$\alpha$ LF. As expected, the overlap phase speeds-up the reionization process; for a star formation time $50,100,200, 300$ Myr, $\langle \chi_{HI} \rangle $ decreases to $\sim 0.24,0.16,4.5\times 10^{-2}, 1.1\times 10^{-2}$, respectively. 

Finally at about 300 Myr (lower center panel), the simulation volume
is almost completely ionized. This configuration corresponds to the 
post-overlap phase which leads to a further acceleration of the 
reionization process. For about 400 Myr of star formation activity, $\langle \chi_{HI} \rangle$ drops to a value of about $4.3 \times 10^{-3}$. This is a consequence of the fact
that, as the IGM becomes more ionized, even the ionizing radiation from the numerous smaller galaxies contributes to the overall radiation field. 
It is very important to point out that, even in this final stage, the
ionization field is highly inhomogeneous, with regions of
higher ionization fraction corresponding to the environment of 
point ionizing sources (green regions in the maps). 
This topology is of fundamental importance in the estimation of the
Ly$\alpha$ LFs, as the Ly$\alpha$ transmissivity associated to each
source depends on the extent of the highly ionized regions as well as on the residual neutral hydrogen within them; a very
tiny fraction of residual neutral hydrogen is sufficient to suppress
the intrinsic Ly$\alpha$ luminosity significantly. Even for an
ionization fraction, $\langle \chi_{HI} \rangle \sim 3.4 \times
10^{-3}$, which results after 500 Myr of star formation, the
transmission value ranges between $42-48$\% for galaxies with
increasing $\dot M_*$; the blue part of the line can be significantly attenuated even by the tiny residual fraction of about $10^{-6}$, close to the source.
% ****************************************************************************                                                                                                                 
\section{The LAE Model}
\label{model}

As mentioned in Sec.~\ref{simulations}, for each galaxy in the SPH
simulation snapshot, we obtain the SFR, $\dot M_*$, the mass weighted
stellar metallicity, $Z_*$, the total gas mass, $M_g$, and the stellar
mass, $M_*$; the stellar age is then calculated as $t_* = M_*/\dot
M_*$, i.e. assuming  a constant SFR. This assumption has been made for
consistency with the RT calculation (see
Sec.~\ref{simulations}). These properties are then used to calculate
the intrinsic Ly$\alpha$ ($L_\alpha^{int}$) and continuum luminosity
($L_c^{int}$), considering the contribution from both stellar sources
and from the cooling of collisionally excited \HI in the ISM of the
galaxy. The intrinsic values of the Ly$\alpha$ and continuum
luminosities so obtained can then be used to calculate the observed
luminosity values. It is comparatively easy to calculate the observed
continuum luminosity, $L_c$, since the wavelength range (1250-1500
\AA) of the continuum luminosity band is chosen so as to be unaffected
by \HI; $L_c$ in fact only depends on the fraction of the continuum photons, $f_c$, that escape the galactic environment undamped by dust, i.e.
\begin{equation}
L_c = L_c^{int} f_c. 
\label{lc_int}
\end{equation}

On the other hand, since Ly$\alpha$ photons have a large absorption
cross-section against \HI, the observed Ly$\alpha$ luminosity,
$L_\alpha$, depends both on: (a) the fraction of Ly$\alpha$ photons,
$f_\alpha$, that escape out of the galaxy into the IGM, undamped by dust and (b) the fraction, $T_\alpha$, of these ``escaped" photons that are transmitted through the IGM, undamped by \HI. The observed Ly$\alpha$ luminosity can therefore be expressed as
\begin{equation}
L_\alpha = L_\alpha^{int} f_\alpha T_\alpha.
\label{la_int}
\end{equation}

All galaxies with $L_\alpha \geq 10^{42.2} {\rm erg \, s^{-1}}$ and an observed equivalent width, $EW = (L_\alpha / L_c) \geq 20$ \AA \,are identified as LAEs and used to construct the cumulative Ly$\alpha$ and UV LFs which can then be compared to the observations. In principle, our results can be compared to both the observed LFs at $z \sim 5.7$ and $6.6$. However, there exists a huge uncertainty between the complete photometric and confirmed spectroscopic sample at $z \sim 6.6$; we limit the comparison between the theoretical model and observations to the data accumulated at $z \sim 5.7$ by Shimasaku et al. (2006), in this work. The calculations of $L_\alpha^{int}$, $L_c^{int}$, $f_\alpha$, $f_c$ and $T_\alpha$ are explained in more detail in what follows.

% *******************
\subsection{The intrinsic luminosity model}
\label{intlum_model}

Star formation produces photons with energies $> 1$~Ryd, which ionize
the \HI in the ISM. Due to the high density of the ISM,
recombinations take place on a short time scale and this produces a
large number of Ly$\alpha$ photons. The stellar luminosity component is calculated by using {\rm STARBURST99} (Leitherer et al. 1999), a population synthesis code, to obtain the spectra and hence, the \HI ionizing photon rate ($Q$) of each galaxy, using a Salpeter IMF, $Z_*$, $t_*$ and $\dot M_*$ as input parameters. Then the intrinsic Ly$\alpha$ luminosity produced by stellar sources $L_\alpha^*$, is calculated as
\begin{equation}
L_\alpha^* = \frac{2}{3} (1-f_{esc}) Q h \nu_\alpha,
\end{equation}
where $f_{esc} = 0.02$, for consistency with the RT calculation (see Sec. \ref{simulations}), $h$ is the Planck constant, $\nu_\alpha$ is the frequency of the Ly$\alpha$ photons and the factor of two-thirds enters assuming case B recombination (Osterbrock 1989). The intrinsic continuum luminosity produced by stellar sources, $L_c^*$, is calculated at 1375 \AA, at the middle of the continuum band between 1250-1500 \AA.

The Ly$\alpha$ and continuum luminosity from the cooling of
collisionally excited \HI in each galaxy ($L_\alpha^g, L_c^g$
respectively) is calculated using the number density of electrons and
\HI, which depend on the temperature distribution of the ISM gas mass
and the gas distribution scale of the galaxy. The temperature
dependence comes from the fact that the recombination coefficient,
which determines the neutral hydrogen fraction in the ISM, decreases sensitively with higher
temperatures. 
The \HI number density is calculated by assuming a primordial ISM composition ($76\%$ hydrogen and $24\%$ helium) and calculating the gas distribution scale assuming the \HI to be concentrated in a radius, $r_g = 4.5\lambda r_{200}$. Here the spin parameter, $\lambda=0.05$ (Ferrara, Pettini \& Shchekinov 2000) and $r_{200}$ is the virial radius, assuming the collapsed region has an overdensity of 200 times the critical density at the redshift considered. The calculation of the electron number density requires a knowledge of the temperature distribution of the IGM gas. However, since we do not have this information from the SPH simulation used in this work, we use the temperature distribution averaged over distinct halo mass ranges shown in Tab. 1 of Dayal et al. (2009b) and again use the gas distribution scale to calculate the electron number density. Complete details of this calculation can be found in Sec.~3.2 of Dayal et al. (2009b). 

% *********************
\subsection{The dust model}
\label{dust_model}

It is important to model the dust in the ISM of galaxies since it absorbs both Ly$\alpha$ and continuum photons, thereby, strongly influencing the observed luminosity values. Dust is produced both by supernovae and evolved stars in a galaxy. However, several authors (Todini \& Ferrara 2001; Dwek et al. 2007) have shown that the contribution of AGB stars can be neglected for $z \gsim 5.7$, since the typical evolutionary time-scale of these stars ($\geq 1$ Gyr) becomes longer than the age of the Universe above that redshift. Hence, assuming SNII to be the primary dust factories at $z \sim 6.1$, we calculate the dust enrichment for each galaxy, taking into account three processes: (a) SNII produce dust in the expanding ejecta; the average dust mass produced per SNII is taken to be ${0.54\, \rm M_\odot}$ (Todini \& Ferrara 2001; Nozawa et al. 2003, 2007; Bianchi \& Schneider 2007), (b) SNII destroy dust in the ISM they shock to velocities $\geq 100 \, {\rm km \, s^{-1}}$, for which we use a destruction efficiency value $\sim 0.12$ (McKee 1989),  and (c) a homogeneous mixture of gas and dust is assimilated into star formation (astration). Once the final dust mass, $M_{dust}(t_*)$, is calculated for each galaxy in the simulation, we can translate this into an optical depth, $\tau_c$, for continuum photons as
\begin{equation}
\tau_c (r_d) = \frac{3\Sigma_{d}}{4 a s},
\label{tau_dust}
\end{equation}
where $\Sigma_d = M_{dust}(t_*) r_d^{-2}$ is the dust surface mass density, $r_d$ is the radius of dust distribution, $a$ and $s$ are the radius and material density of graphite/carbonaceous grains, respectively ($a=0.05 \mu m$, $s = 2.25\, {\rm g\, cm^{-3}}$; Todini \& Ferrara 2001; Nozawa et al. 2003). This can be used to obtain the escape fraction of continuum photons from the galaxy as $f_c = e^{-\tau_c}$. The determination of $r_d$ follows in Sec.~\ref{reio effects} and complete details of the calculations mentioned here can be found in Dayal et al. (2009b).

Different extinction curves, including that for the Milky Way, Small Magellanic Cloud and for supernovae give different relations between the extinction of the Ly$\alpha$ and continuum photons, for homogeneously distributed dust. In this spirit, and to get a hint of the dust inhomogeneity, we combine $f_\alpha$ and $f_c$, and use the escape fraction of Ly$\alpha$ photons relative to the continuum ones, $f_\alpha/f_c$, to be the free parameter to calculate $L_\alpha$ in our model. A more detailed explanation of the determination of $f_c$ and $f_\alpha$, required to reproduce the observations, follows later in Sec.~\ref{reio effects}.

%In the same spirit, and following earlier works on LAEs (Finkelstein et al. 2009, Kobayashi et al. 2009, Dayal et al. 2008, 2009ab), we combine $f_\alpha$ and $f_c$ and use the escape fraction of Ly$\alpha$ photons relative to the continuum ones, $f_\alpha/f_c$, to be the free parameter to calculate $L_\alpha$ in our model. A more detailed explanation of the determination of $f_c$ and $f_\alpha$, required to reproduce the observations, follow later in Sec.~\ref{reio effects}.

% *************************
\subsection{The IGM transmission model}
\label{ta_model}

The calculation of $T_\alpha$ is now explained in better detail. If
$z_{em}$ and $z_{obs}$ are the redshifts of the emitter and the observer respectively, we calculate the total optical depth ($\tau_\alpha$) to the Ly$\alpha$ photons along a line of sight (LOS) as 
\begin{equation}
\tau_\alpha({\rm v}) = \int_{z_{em}}^{z_{obs}} \sigma_0 \phi({\rm v}) n_{HI}(z) \frac{dl}{dz} dz ,
\end{equation}
where ${\rm v} = (\lambda - \lambda_\alpha)[\lambda_\alpha c]^{-1}$ is the rest-frame velocity of a photon with wavelength $\lambda$, relative to the line centre (rest-frame wavelength $\lambda_\alpha = 1216$ \AA, velocity ${\rm v}_\alpha =0$), $\phi$ is the Voigt profile, $n_{HI}$ is the number density of \HI along the LOS, $dl/dz = c[H(z) (1+z)]^{-1}$ and $c$ represents the light speed. We use $\sigma_0 = \pi e^2 f[m_e c]^{-1}$, where $e$, $m_e$ are the electron charge and mass respectively and $f$ is the oscillator strength (0.4162). 

For regions of low \HI density, the natural line broadening is not very important and the Voigt profile can be approximated by the Gaussian core: 
\begin{equation} 
\phi({\rm v}_i) \equiv \phi_{gauss} = \frac{\lambda_\alpha}{\sqrt{\pi} b} e^{-(\frac{{\rm v}_i + {\rm v}_p-{\rm v}_\alpha}{b})^2},
\label{gauss} 
\end{equation} 
where ${\rm v}_i$ is the velocity of a photon of initial velocity ${\rm v}$ at a redshift $z_i$ along the LOS, ${\rm v}_p$ is the peculiar velocity at $z_i$ and  ${\rm v}_\alpha$ is the velocity of the Ly$\alpha$ photons at $z_i$, which is 0 in this expression. The Doppler width parameter is expressed as $b=\sqrt {2kT/m_H}$, where $m_H$ is the hydrogen mass, $k$ is the Boltzmann constant and $T$ is the IGM temperature. While we use a $T=10^4 K$ for each cell with an ionization fraction $(1-\chi_{HI})\geq 0.1$, we use $T=20$ K if the IGM is pristine, i.e. has an ionization fraction $(1-\chi_{HI})\leq 0.1$. \footnote{We find that even for an average neutral hydrogen fraction as high as, $\langle \chi_{HI} \rangle \sim 0.295$, only 8 cells out of $256^3$ have $(1-\chi_{HI})\leq 0.1$, with the complete RT simulation as described in Sec.~\ref{simulations}. Hence, to simplify the calculations, we use $T =10^4$ K for all the cells for all the {\tt CRASH} outputs.}

In more dense regions the Lorentzian damping wing of the Voigt profile becomes important. According to Peebles (1993), this can be approximated as  
\begin{equation} 
\phi_{lorentz}({\rm v}_i) = \frac{R_\alpha \lambda_\alpha}{\pi[({\rm v}_i + {\rm v}_p-{\rm v}_\alpha)^2 + R_\alpha^2]} 
\label{lorentz}
\end{equation} 
where $R_\alpha = \Lambda \lambda_\alpha [4 \pi]^{-1}$ and $\Lambda= 6.25\times 10^8$~s$^{-1}$ is the decay constant for the Ly$\alpha$ resonance. Although computationally more expensive than the above approximations, using the Voigt profile to compute the absorption cross-section gives precise results, and therefore we have implemented it in our code to obtain all the results presented below.  

The total Ly$\alpha$ optical depth is calculated for each galaxy identified in the {\tt CRASH} 
outputs corresponding to $\langle \chi_{HI} \rangle \sim 0.29, 0.24,
0.16, 4.5\times 10^{-2}, 1.1 \times 10^{-2}, 4.3\times 10^{-3}, 3.4
\times 10^{-3}$ as mentioned in Sec.~\ref{simulations}. The
transmission along any LOS is calculated as $T_\alpha =
e^{-\tau_\alpha}$, integrating from the position of each galaxy to the edge of the box. To get a statistical estimate of the transmission, we
construct 6 LOS starting from the position of each galaxy, one along
each of the box axes, for each of the mentioned ionization states. The
average value of the transmission for each galaxy is then obtained by
averaging the transmission over the 6 LOS. This is done for all the
ionization state configurations resulting from the RT calculation.

Once the above model is in place, the only {\it two free parameters} of our model to match the theoretical Ly$\alpha$ and UV LFs to the observed ones are: the dust distribution radius, $r_d$, and the relative escape fraction of Ly$\alpha$ photons as compared to the continuum photons, $f_\alpha/f_c$. Both these parameters remain quite poorly understood due to a lack of observational data about the dust distribution/topology in high-redshift galaxies; they must therefore be inferred by comparing the theoretical LFs to the observed ones. Details on how these parameters are determined follow in Sec.~\ref{reio effects}.

% ****************************************************************************************
\section{LAE visibility during reionization}
\label{reio effects}

Once the combined SPH+RT calculations are carried out and the LAE
model implemented, we are in a position to quantify the importance of
reionization, peculiar velocities and the dust enrichment on the
observed LFs. Physically, the reionization process leads to a decrease in $\langle \chi_{HI} \rangle$, thereby increasing $T_\alpha$; on the
other hand, peculiar velocities caused by galactic scale outflows (inflows), redshift (blueshift) the Ly$\alpha$ photons, thereby leading to a higher (lower) value of $T_\alpha$. A handle on the dust enrichment is necessary since dust grains absorb both Ly$\alpha$ and continuum photons, thereby affecting their escape fractions from the galaxy.   

\begin{figure} 
%  \vspace*{10pt} 
  \center{\includegraphics[scale=0.45] {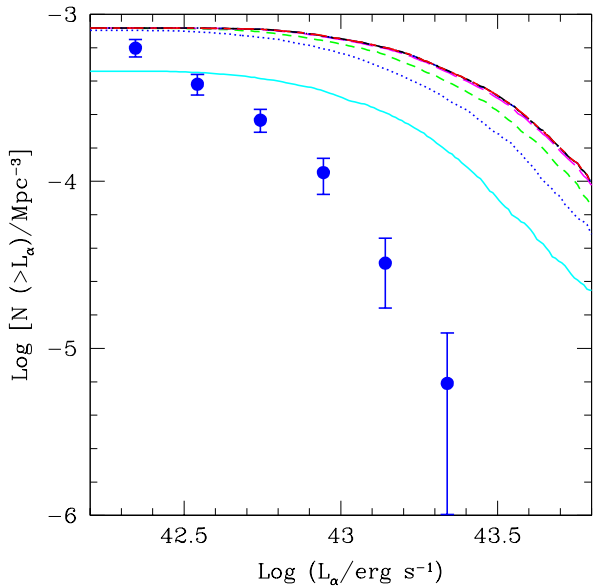}} 
 \caption{Cumulative Ly$\alpha$ LF for galaxies identified as LAEs from the SPH simulation snapshot including the full RT calculation and density fields but ignoring velocity fields (${\rm v}_p=0$) and dust ($f_\alpha=f_c = 1$). The lines from bottom to top correspond to {\tt CRASH} outputs with decreasing values of $\langle \chi_{HI} \rangle$ such that: $\langle \chi_{HI} \rangle \sim$ = 0.29 (solid), 0.24 (dotted), 0.16 (short dashed), $4.5 \times 10^{-2}$ (long dashed), $1.1 \times 10^{-2}$ (dot-short dashed), $4.3 \times 10^{-3}$ (dot-long dashed) and $3.4 \times 10^{-3}$ (short-long dashed). Points show the observed LF at $z\sim 5.7$ (Shimasaku et al. 2006). }
\label{lya_novel_nodust}
\end{figure} 
 
We begin our study by ignoring the effects of the peculiar velocities and assuming all the galaxies to be dust free, to  quantify how each of these two parameters shapes the observed Ly$\alpha$ and UV LFs. The former assumption means that ${\rm v}_p = 0$ in Eqs. \ref{gauss}, \ref{lorentz}; the latter implies that all photons produced inside the galaxy escape into the IGM, i.e. $f_c = f_\alpha = 1$ in Eqs. \ref{lc_int}, \ref{la_int}. We then use the prescriptions detailed in Sec.~\ref{model} to identify the galaxies that would be visible as LAEs in each {\tt CRASH} output and build their Ly$\alpha$ LF, as shown in Fig.~\ref{lya_novel_nodust}. 

Since both peculiar velocities and dust are neglected, in this case,
while the UV LF is simply the intrinsic LF, the Ly$\alpha$ LF is
shaped solely by the transmission through the IGM. As mentioned in
Sec.~\ref{simulations}, after a star formation time scale as short as
10 Myr ($\langle \chi_{HI} \rangle \sim 0.29$), the galaxies in the
simulation snapshot are able to form \HII regions, with the region
size and the ionization fraction inside it increasing with
$\dot M_*$; this results in a $T_\alpha$ which increases with $\dot
M_*$. As the galaxies continue to form stars, the sizes of the \HII
regions built by each source increase with time, leading $\langle
\chi_{HI} \rangle$ to decrease to $0.24,0.16$ at  $t=50, 100$ Myr
respectively. This leads to an increase in the transmission of the red
part of the Ly$\alpha$ line for all the sources, in turn yielding a
corresponding increase of the Ly$\alpha$ luminosity of each source, as
depicted in Fig.~\ref{lya_novel_nodust}. However, after about 200 Myr,
$\langle \chi_{HI} \rangle$ reduces to $\sim 0.04$; the \HII regions
built by each source are large enough so that almost all of the red
part of the Ly$\alpha$ line is transmitted and the value of $T_\alpha$
saturates for all LAEs. This results in very similar LFs for $\langle \chi_{HI} \rangle \leq 0.04 $, as seen from the same figure. 

\begin{figure} 
%  \vspace*{10pt} 
  \center{\includegraphics[scale=0.45]{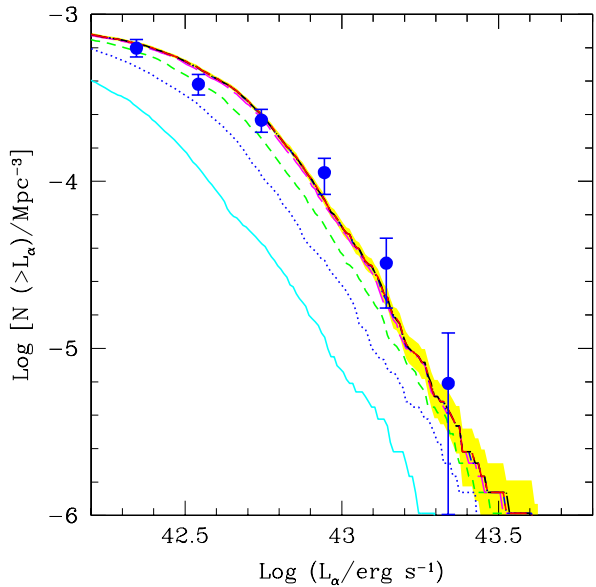}} 
  \caption{Cumulative Ly$\alpha$ LF for galaxies identified as LAEs from the SPH simulation snapshot including the full RT calculation, density fields and dust but ignoring velocity fields (${\rm v}_p=0$), corresponding to models ${\rm S1-S7}$ in Tab. \ref{tab1}. The lines from bottom to top correspond to {\tt CRASH} outputs with decreasing values of $\langle \chi_{HI} \rangle$ such that: $\langle \chi_{HI} \rangle \sim$  0.29 (solid), 0.24 (dotted), 0.16 (short dashed), $4.5 \times 10^{-2}$ (long dashed), $1.1 \times 10^{-2}$ (dot-short dashed), $4.3 \times 10^{-3}$ (dot-long dashed) and $3.4 \times 10^{-3}$ (short-long dashed). The shaded region shows the poissonian error corresponding to $\langle \chi_{HI} \rangle \sim 3.4 \times 10^{-3}$, i.e., model ${\rm S7}$ (Tab. \ref{tab1}) and points show the observed LF at $z\sim 5.7$ (Shimasaku et al. 2006).}
\label{lya_novel_incdust} 
\end{figure}

\begin{figure*} 
%  \vspace*{10pt} 
  \center{\includegraphics[scale=1.0]{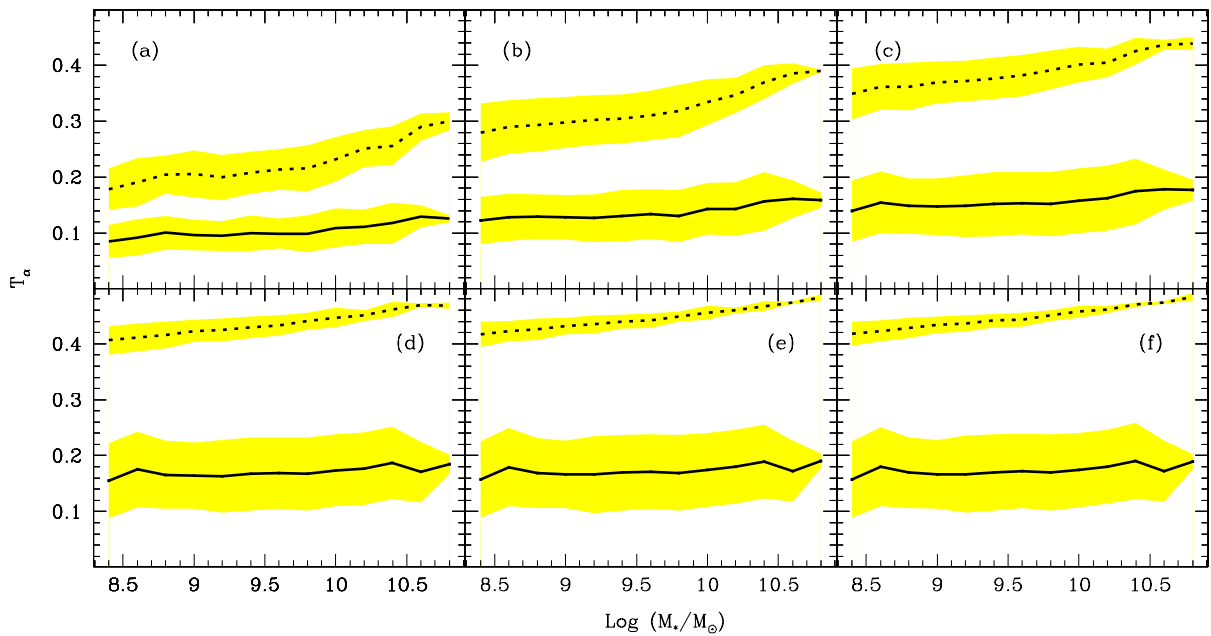}} 
  \caption{$T_\alpha$ as a function of $M_*$ for galaxies identified as LAEs from the simulation snapshot including the full RT calculation, density fields, dust and ignoring (including) velocity fields shown by dotted (solid) lines in each panel. The solid/dotted lines in each panel correspond to the following models given in Tab. \ref{tab1}: (a) ${\rm M1}/{\rm S1}$ ($\langle \chi_{HI} \rangle  \sim 0.29$), (b) ${\rm M2}/{\rm S2}$ (0.24), (c) ${\rm M3}/{\rm S3}$ (0.16), (d) ${\rm M4}/{\rm S4}$ ($4.5 \times 10^{-2}$), (e) ${\rm M5}/{\rm S5}$ ($1.1 \times 10^{-2}$) and (f) ${\rm M7}/{\rm S7}$ ($3.4 \times 10^{-3}$). In each panel, the stellar mass bins span 0.2 dex and the shaded regions represent the $1\sigma$ error bars in each mass bin. }
\label{fractx_ms} 
\end{figure*}

The key point here is that, if peculiar velocities and dust effects are neglected, at no stage of the reionization history, either the slope or the amplitude of the observed Ly$\alpha$ LF can be reproduced; analogous problems arise also when the UV LFs are considered.

%{\tt Fig.~4: why do you plot errors with shade and bars; what are the errors ? get rid of the UVB photorate curves; why is the Gamma so low ? And by UVB do you mean the sum of galaxy+UVB ? you often confused the  two - CHECK}.

The above discussion implies the need of one or more physical effects attenuating the Ly$\alpha$ and continuum photons, the most obvious of which is the presence of dust in the ISM of these galaxies, which would absorb both Ly$\alpha$ and continuum photons, simultaneously reducing $f_\alpha$ and $f_c$. We then include the dust model described in Sec.~\ref{dust_model} into our calculations. Since the UV is unaffected by the \HI in the IGM as mentioned in Sec.~\ref{model}, the same value of $r_d = 0.48 r_g$ ($r_g$ is the ISM gas distribution scale, Sec.~\ref{dust_model}) reproduces the UV LF for all of the ionization states of the IGM, ranging from $\langle \chi_{HI} \rangle \sim 0.29$ to $3.4 \times 10^{-3}$. This assumption of the dust distribution scale leads to an average escape fraction of continuum photons, $f_c \sim 0.12$ (refer Sec.~\ref{dust_model} for details) for the galaxies we identify as LAEs.

We now momentarily digress to discuss the effect of reionization on $T_\alpha$ (defined by Eq. \ref{la_int}) for different stellar mass ranges. However, since $\dot M_* $ scales with $M_*$ (details in Sec. \ref{phy prop}), we discuss $T_\alpha$ in terms of $\dot M_*$. As mentioned in Sec.~\ref{simulations}, we initialize the {\tt CRASH} runs with $\langle \chi_{HI} \rangle \sim 0.3$. In a star formation timescale of 10 Myr, by virtue of the \HII regions that already start growing, the total photoionization rate (sum of contributions from the UVB and all the galaxies in the simulation) has a value of $\Gamma_T$\footnote{ $\Gamma_T$, the total photoionization rate, is calculated assuming ionization-recombination equilibrium over average values of density and ionization fraction in the simulation volume, so that $$\Gamma_T = \frac{(1-\langle \chi_{HI} \rangle)^2 \langle n_H \rangle \alpha_B}{\langle \chi_{HI} \rangle}.$$ Here, $\langle n_H \rangle$ is the average hydrogen number density in the simulation volume and $\alpha_B$ is the case B recombination co-efficient. } $= 2.8 \times 10^{-17} {\rm s^{-1}}$ and $\langle \chi_{HI} \rangle$  decreases slightly to 0.295. At this point, due to their smaller \HI ionizing photon output, galaxies with $\dot M_* \leq 25 \, {\rm M_\odot \, yr^{-1}}$,  have $T_\alpha \sim 0.2$; galaxies with larger $\dot M_*$, have $T_\alpha \sim 0.3$, as shown in Panel (a) of Fig.~\ref{fractx_ms}. As the star formation continues, for 50 (100) Myr of star formation, the \HII region sizes increase and $\Gamma_T$ increases slightly to $\sim 4.0 \times 10^{-17}$ $(7.7 \times 10^{-17}) \, {\rm s^{-1}}$; $T_\alpha$ increases and ranges between $0.28-0.4$ (0.34-0.44) for $\dot M_* \sim 8-200 \, {\rm M_\odot \, yr^{-1}}$, Panel b (c), Fig.~\ref{fractx_ms}. Finally for $\Gamma_T \sim 3.3 \times 10^{-16} \, {\rm s^{-1}}$, corresponding to $\langle \chi_{HI} \rangle \sim 0.04$, the transmission settles to $T_\alpha \sim 0.4-0.48$ for $\dot M_* \sim 8-200 \, {\rm M_\odot \, yr^{-1}}$ (Panels d-f); in about 200 Myr from the ignition of star formation, the Str\"omgren spheres built by these LAEs are large enough so that the redshifted Ly$\alpha$ photons are no longer affected by the \HI outside this region. However, the residual \HI inside this ionized region leads to an absorption of the photons blueward of the Ly$\alpha$ line, and hence, about half of the line is transmitted. The values of $\Gamma_T$, $\langle \chi_{HI} \rangle$ and the $T_\alpha$ values averaged for all LAEs in each {\tt CRASH} output are shown in Tab. \ref{tab1}.

\begin{figure} 
%  \vspace*{10pt} 
  \center{\includegraphics[scale=0.45]{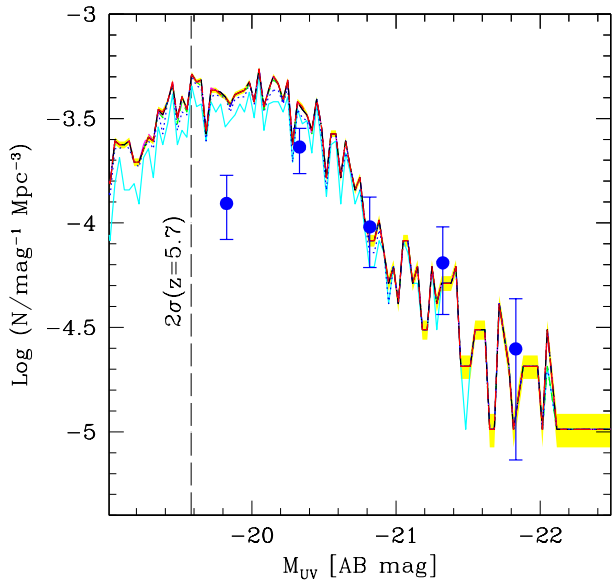}} 
  \caption{UV LF for galaxies identified as LAEs from the simulation snapshot including the full RT calculation, density fields, velocity fields and dust, corresponding to models ${\rm M1}$ to ${\rm M7}$ in Tab. \ref{tab1}. The lines from bottom to top correspond to decreasing values of $\langle \chi_{HI} \rangle$ such that: $\langle \chi_{HI} \rangle \sim$ 0.29 (solid), 0.24 (dotted), 0.16 (short dashed), $4.5 \times 10^{-2}$ (long dashed), $1.1 \times 10^{-2}$ (dot-short dashed), $4.3 \times 10^{-3}$ (dot-long dashed) and $3.4 \times 10^{-3}$ (short-long dashed). The shaded region shows the poissonian error corresponding to $\langle \chi_{HI} \rangle \sim 3.4 \times 10^{-3}$, i.e., model ${\rm M7}$ (Tab .\ref{tab1}) and points show the observed UV LF at $z\sim 5.7$ (Shimasaku et al. 2006).}
\label{uv_incvel_incdust} 
\end{figure} 

Using the above $T_\alpha$ values and dust model fixed by the UV LF, the only free parameter we are left with, to match the theoretical and observed Ly$\alpha$ LFs is $f_\alpha/f_c$; this only scales the Ly$\alpha$ LF without affecting its shape. As mentioned above, the $T_\alpha$ value for galaxies identified as LAEs settles for $\langle \chi_{HI} \rangle \leq 0.04$, which leads to a corresponding saturation in the Ly$\alpha$ LF. We find that $f_\alpha/f_c \sim 1.3$ reproduces the slope and the magnitude of the observed Ly$\alpha$ LF quite well for all the {\tt CRASH} outputs where $\langle \chi_{HI} \rangle \leq 0.04$, as shown in Fig.~\ref{lya_novel_incdust}. However, due to decreasing values of $T_\alpha$, the Ly$\alpha$ LF is progressively under-estimated for increasing values of $\langle \chi_{HI} \rangle$, as seen from the same figure.

\begin{figure} 
%  \vspace*{10pt} 
  \center{\includegraphics[scale=0.45]{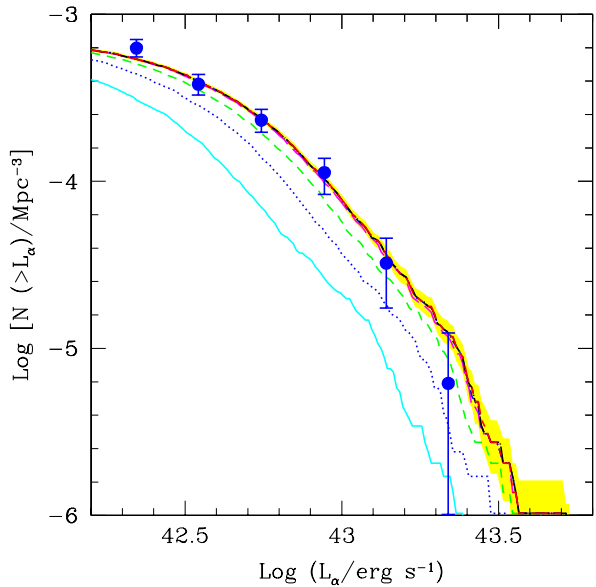}} 
  \caption{Cumulative Ly$\alpha$ LF for galaxies identified as LAEs from the simulation snapshot including the full RT calculation, density fields, velocity fields and dust, corresponding to models ${\rm M1}$ to ${\rm M7}$ in Tab. \ref{tab1}. The lines from bottom to top correspond to decreasing values of $\langle \chi_{HI} \rangle$ such that: $\langle \chi_{HI} \rangle \sim$ 0.29 (solid), 0.24 (dotted), 0.16 (short dashed), $4.5 \times 10^{-2}$ (long dashed), $1.1 \times 10^{-2}$ (dot-short dashed), $4.3 \times 10^{-3}$ (dot-long dashed) and $3.4 \times 10^{-3}$ (short-long dashed). The shaded region shows the poissonian error corresponding to $\langle \chi_{HI} \rangle \sim 3.4 \times 10^{-3}$, i.e. model ${\rm M7}$ (Tab. \ref{tab1}) and points show the observed LF at $z\sim 5.7$ (Shimasaku et al. 2006).}
\label{lya_incvel_incdust} 
\end{figure}

Once the above framework is in place, we include and study the effect of the final component that can affect the observed luminosity, the presence of peculiar velocities. Once this is done by consistently deriving the peculiar velocity field from the simulations, we again reproduce both the magnitude and slope of the observed UV LF as shown in Fig.~\ref{uv_incvel_incdust}. Of course this does not come as a surprise since velocity fields do not affect UV photons.
However, galactic scale outflows (inflows) from a galaxy, redshift (blueshift) the Ly$\alpha$ photons, thereby leading to a higher (lower) $T_\alpha$ value. It is quite interesting to see that we can again match the theoretical Ly$\alpha$ LFs to the observed ones by a simple scaling between $f_\alpha$ and $f_c$ for $\langle \chi_{HI} \rangle \leq 0.04$, as seen from Fig.~\ref{lya_incvel_incdust}. However, we require a much higher value of $f_\alpha/f_c \sim 3.7$ (models $M4-M7$, Tab. \ref{tab1}), compared to the value of $1.3$ excluding velocity fields (models $S4-S7$, Tab. \ref{tab1}). Again, $T_\alpha$, and hence the Ly$\alpha$ LF, get progressively more damped with an increase in $\langle \chi_{HI} \rangle$, as seen from Fig.~\ref{lya_incvel_incdust}, (models $M3-M1$, Tab. \ref{tab1}).

We now discuss the reason for the higher $f_\alpha/f_c \sim 3.7$ value required to reproduce the Ly$\alpha$ LF when velocity fields are considered, as compared to the ratio of $1.3$, when they are not, for $\langle \chi_{HI} \rangle \leq 0.04$.
The galaxies we identify as LAEs have halo masses $M_h \sim
10^{10.4-12} \, M_\odot$, which correspond to $\geq 2\sigma$
fluctuations at $z \sim 6.1$. These LAEs are therefore subject to
strong inflows since they lie in dense regions. As these inflows
blue-shift the Ly$\alpha$ photons, even photons in the red part of the
line are attenuated, thereby reducing $T_\alpha$. This can be seen
clearly from Fig.~\ref{fractx_ms}, where the solid lines, which
represent the outcomes from the models including velocities, are
always significantly below the dotted ones, which represent the model
with no velocity field included. When velocity fields are included we find
$T_\alpha \sim 0.08-0.12$ ($0.16-0.18$) for $\langle \chi_{HI} \rangle \sim 0.29$ ($3.4\times
10^{-3}$) for $\dot M_* \sim 8-200 \, {\rm M_\odot \, yr^{-1}}$. Averaging over all the LAEs for $\langle \chi_{HI}\rangle \sim 3.4\times 10^{-3}$, we find the value of $T_\alpha$ $\sim 0.17$ ($0.44$) including (excluding) the effects of peculiar velocity fields. This requires that when peculiar velocities are included, a correspondingly larger fraction ($0.44/0.17 = 2.6$) of Ly$\alpha$ photons must escape the galaxy, undamped by dust to bring the observed luminosity up to the levels it would reach in the absence of these inflows. Although many galaxies do show outflows, powered by supernova explosions, these dominate at very small scales ($\leq 170$ physical Mpc). However, the small redshift boost imparted by these is negligible compared to the blue-shifting of the Ly$\alpha$ line because of large scale inflows; eventually, it is these dominant inflows that determine the value of $T_\alpha$.

Further, including velocity fields changes the slope of the LF; when
velocity fields are not included, $T_\alpha$ basically scales with $\dot M_*$; when these are included, $T_\alpha$ is the most damped for the
largest masses, since these see the strongest inflow velocities by
virtue of their largest potential wells. 
This can be seen clearly from Fig.~\ref{fractx_ms} where the slope of
$T_\alpha$ is visibly shallower for the solid curves (including
velocities) than for the dotted ones (neglecting velocities). This has
the effect of {\it flattening} the slope of the Ly$\alpha$ LF as
can be seen in Fig.~\ref{lya_incvel_incdust}). 

\begin{figure} 
%  \vspace*{10pt} 
  \center{\includegraphics[scale=0.5]{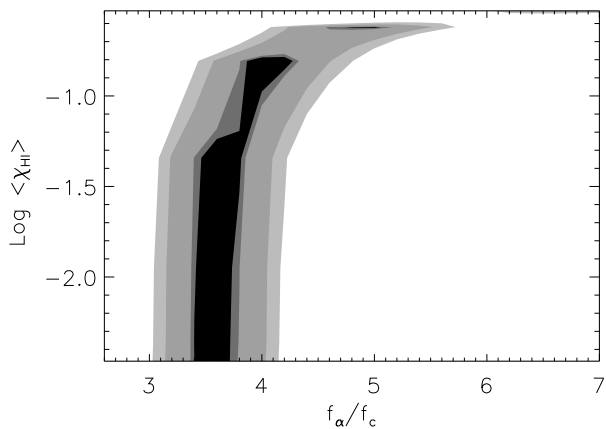}} 
 \caption{The $1-5\sigma$ probability contours (black to light gray respectively) for combinations of $\langle \chi_{HI} \rangle$ and $f_\alpha/f_c$ that fit the observed Ly$\alpha$ LF.  Within a $1\sigma$ ($5\sigma$) error, we can not distinguish an IGM with $\langle \chi_{HI} \rangle \sim 3.4 \times 10^{-3}$ from one where $\langle \chi_{HI} \rangle \sim 0.16$ (0.24) for $f_\alpha/f_c \sim 3.4-4.1$ ($3.0-5.7$). This implies a {\it degeneracy} between the ionization state of the IGM and dust clumping (or grain properties) inside high-redshift galaxies; the ionization state of the IGM cannot be tightly constrained unless the relative escape fraction of Ly$\alpha$ compared to the continuum photons is reasonably well understood. Refer Sec.~\ref{reio effects} for details. }
\label{prob_contours}
\end{figure} 

\begin{table*} 
\begin{center} 
\caption {The model designation (col 1), the star formation timescale to ionize the IGM (col 2), the features included in the model (VF stands for peculiar velocity fields) (col 3), the average neutral hydrogen fraction (col 4), the value of the total photoionization rate (sum of the contribution from the UVB and from all the galaxies in the snapshot) corresponding to this neutral hydrogen fraction (col 5), the average transmission for all the galaxies identified as LAEs (col 6), the average escape fraction of continuum photons for LAEs for the model presented in Sec.~\ref{dust_model} (col 7), the average color excess value corresponding to this continuum escape fraction using the supernova extinction curve (col 8) and the relative escape fraction of Ly$\alpha$ and continuum photons to best fit the observations for $\langle \chi_{HI} \rangle \leq 0.04$ (col 9).}
\begin{tabular}{|c|c|c|c|c|c|c|c|c} 
\hline 
${\rm Model}$ & $T$ & ${\rm Features}$ & $\langle \chi_{HI} \rangle$ & $\Gamma_T$ & $\langle T_\alpha \rangle$ &
$\langle f_c \rangle $ & $\langle E(B-V) \rangle $ & $f_\alpha/f_c$ \\
$$ & $[{\rm Myr}]$ & $$ & $$ & $[10^{-14}\, {\rm s^{-1}}]$ & $$ & $$ & $$ & $$ \\
\hline
${\rm S1}$ & $10$ & ${\rm RT + dust}$ & $0.295$ & $2.8 \times 10^{-3}$  & $0.21$ & $0.12$ & $0.2$ & $1.3$\\
${\rm S2}$ & $50$ & ${\rm RT + dust}$ & $0.240$ & $4.0 \times 10^{-3}$  &  $0.30$ & $0.12$ & $0.2$ & $1.3$\\
${\rm S3}$ & $100$ & ${\rm RT + dust}$ & $0.157$ & $7.6 \times 10^{-3}$   & $0.37$ & $0.12$ & $0.2$ & $1.3$\\
${\rm S4}$ & $200$ & ${\rm RT + dust}$ & $4.55\times 10^{-2}$ & $3.3 \times 10^{-2}$  & $0.42$ &  $0.12$ & $0.2$  &$1.3$\\
${\rm S5}$ & $300$ & ${\rm RT + dust}$ & $1.13 \times 10^{-2}$ & $0.14$  & $0.43$ & $0.12$ & $0.2$ &$1.3$\\
${\rm S6}$ & $400$ & ${\rm RT + dust}$ & $4.33 \times 10^{-3}$ & $0.33$ & $0.44$ & $0.12$ & $0.2$& $1.3$\\
${\rm S7}$ & $500$ & ${\rm RT + dust}$ & $3.43 \times 10^{-3}$ & $0.48$   & $0.44$ & $0.12$ & $0.2$ & $1.3$\\

$$ & $$ & $$ & $$ & $$  & $$ & $$ & $$ & $$ \\

${\rm M1}$ & $10$ & ${\rm RT + VF + dust}$ & $0.295$ & $2.8 \times 10^{-3}$   & $0.10$ & $0.12$ & $0.2$ & $3.7$\\
${\rm M2}$ & $50$ & ${\rm RT + VF +dust}$ & $0.240$ & $4.0 \times 10^{-3}$   & $0.13$ & $0.12$ & $0.2$ & $3.7$\\
${\rm M3}$ & $100$ & ${\rm RT + VF +dust}$ & $0.157$ & $7.6 \times 10^{-3}$  & $0.15$ & $0.12$ & $0.2$ &$3.7$\\
${\rm M4}$ & $200$ & ${\rm RT + VF +dust}$ & $4.55 \times 10^{-2}$ & $3.3 \times 10^{-2}$  & $0.16$ & $0.12$ & $0.2$ &$3.7$\\
${\rm M5}$ & $300$ & ${\rm RT + VF +dust}$ & $1.13 \times 10^{-2}$ & $0.14$  & $0.17$ & $0.12$ & $0.2$ & $3.7$\\
${\rm M6}$ & $400$ & ${\rm RT + VF +dust}$ & $4.33 \times 10^{-3}$ & $0.33$ & $0.17$ & $0.12$  & $0.2$ & $3.7$\\
${\rm M7}$ & $500$ & ${\rm RT + VF +dust}$ & $3.43 \times 10^{-3}$ & $0.48$ & $0.17$ & $0.12$ & $0.2$  &  $3.7$\\
\hline
\label{tab1} 
\end{tabular} 
\end{center}
\end{table*}

As an important result, the above analysis shows that there exists a degeneracy between the ionization state of the IGM and dust clumping (or grain properties) inside high-redshift galaxies, i.e a high (low) $T_\alpha$ can be compensated by a low (high) $f_\alpha$. This is shown in Fig.~\ref{prob_contours}, where we find that within a $1\sigma$ error, for $f_\alpha/f_c \sim 3.4-4.1$, we can not distinguish an IGM with $\langle \chi_{HI} \rangle \sim 3.4 \times 10^{-3}$ from one where $\langle \chi_{HI} \rangle \sim 0.16$. Within the area under the $5\sigma$ error, $f_\alpha/f_c \sim 4.1- 5.7$ can also fit the Ly$\alpha$ LF for $\langle \chi_{HI} \rangle \sim 0.24$. This leads to the very interesting conclusion that the ionization state of the IGM cannot be tightly constrained unless the relative escape fraction of Ly$\alpha$ compared to the continuum photons is reasonably well understood.

We now discuss the dusty nature of the LAEs we identify from the SPH simulation. We require that for the complete  LAE model ($M1-M7$, Tab. \ref{tab1}), the value of $f_\alpha/f_c$ must range between $3.4-4.1$ ($3-5.7$) for an average neutral hydrogen fraction of $\langle \chi_{HI} \rangle \leq 0.16$ ($\leq 0.24$). However, no single extinction curve gives a value of $f_\alpha/f_c >1$. One of the simplest ways of explaining this large relative escape fraction is to invoke the multiphase ISM model as proposed by Neufeld (1991), wherein the ISM is multiphase and consists of a warm gas with cold dust clumps embedded in it. This inhomogeneity of the dust distribution can then lead to a larger attenuation of the continuum photons relative to the Ly$\alpha$.

\begin{figure*} 
%  \vspace*{10pt} 
  \center{\includegraphics[scale=1.05]{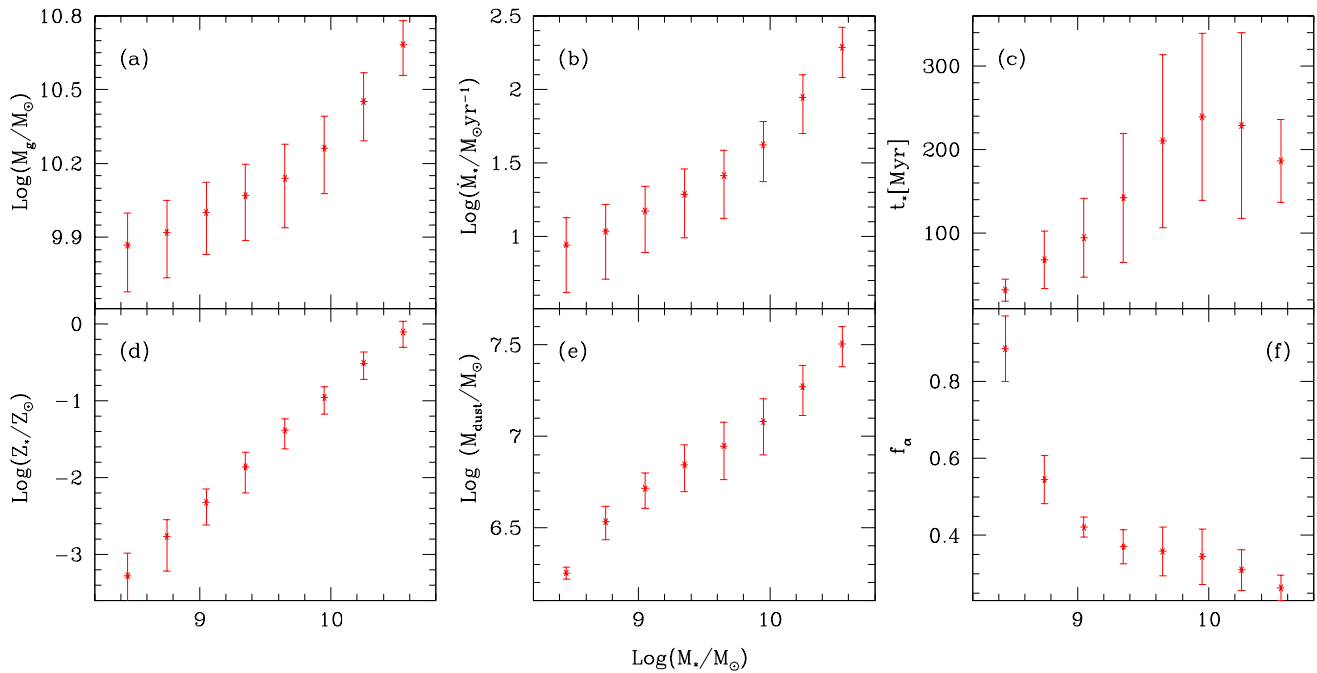}} %phy_prop_laes_5d8
  \caption{Physical properties of the galaxies identified as LAEs using model ${\rm M7}$ in Tab. \ref{tab1}. This model includes the full RT calculation, density, velocity fields, dust and has $\langle \chi_{HI} \rangle \sim 3.4 \times 10^{-3}$. The panels show (a) the gas mass, $M_g$ (b) SFR, $\dot M_*$, (c) mass weighted stellar age, $t_*$ and (d) the mass weighted stellar metallicity, $Z_*$, (e) the total dust mass, $M_{dust}$ and, (f) the escape fraction of Ly$\alpha$ photons, $f_\alpha$,  plotted as a function of the stellar mass, $M_*$. The values of each quantity are shown averaged over $M_*$ bins spanning 0.2 dex, with the error bars representing the $1\sigma$ error in each bin. }
\label{phy_prop} 
\end{figure*}      

We also translate $f_c$ into the color excess, $E(B-V)$, for models $M1-M7$ (Tab. \ref{tab1}) and compare this value to other high redshift LAE observations. At high redshifts, the observed properties of the most distant quasars (Maiolino et al. 2006) and gamma-ray bursts (Stratta et al. 2007) can be successfully interpreted using a SN extinction curve (Todini \& Ferrara 2001; Bianchi \& Schneider 2007). Using the same curve, we find the average value of the color excess, $E(B-V)= 0.2$, corresponding to an average $f_c=0.12$. This inferred color excess value is in good agreement with recent experimental determinations: by fitting the SEDs of 3 LAEs at $z = 5.7$, Lai et al. (2007) have inferred $E(B-V) < 0.225-0.425$; in a sample of 12 LAEs at $z = 4.5$, Finkelstein et al. (2009) have found $E(B-V) = 0.035-0.316$; finally, Pirzkal et al. (2007) have found $E(B-V)=0.025-0.3$ for 3 galaxies at $z = 4-5.76$.

We now summarize our two main results: (a) we find that the Ly$\alpha$ LF can be well reproduced (to within a $5\sigma$ error) by an average neutral hydrogen fraction as high as 0.24 (an almost neutral IGM), to a value as low as $3.4 \times 10^{-3}$, corresponding to an ionized IGM, provided that the increase in the transmission is compensated by a decrease in the Ly$\alpha$ escape fraction from the galaxy, (b) we find that to reproduce the Ly$\alpha$ LF, for any ionization state of the IGM, we require $f_\alpha/f_c>1$, a value that cannot be obtained using any existing extinction curve; this raises the need to invoke a multiphase ISM model, in which dust clumps are embedded in a highly ionized ISM, to facilitate the Ly$\alpha$ photon escape relative to that of continuum photons.

% ********************************************************************************************
\section{Physical properties of LAEs}
\label{phy prop}

We are now in a position to discuss the physical properties of the galaxies we identify as LAEs, including the stellar and gas mass, metallicity, stellar ages, dust mass and escape fraction of Ly$\alpha$ photons from the ISM. As mentioned in Sec.~\ref{reio effects}, while the UV LF is independent of the ionization state of the IGM, the Ly$\alpha$ LF can be well reproduced if a low (high) Ly$\alpha$ escape fraction from the galaxy is compensated by a high (low) transmission through the IGM. Hence, by scaling up $f_\alpha/f_c$ for increasing values of $\langle \chi_{HI} \rangle > 0.04$ (models M1-M3, Tab. \ref{tab1}), we would broadly always identify the same galaxy population as LAEs. We now show the physical properties for the LAEs identified using model M7 (Tab. \ref{tab1}), which includes the full RT calculation, density/velocity fields and dust, with $f_\alpha/f_c = 3.7$. 

%{\tt How large are the stellar mass (horizontal) bins in Fig.~8 ? Add in number caption or  h-error bars in the Figure.}
We find that there is the direct correlation between $M_g$ and $M_*$;
LAEs with a larger $M_*$ are also more gas rich. However, the ratio
$M_g/M_*$ is the largest (smallest) for LAEs with the smallest
(largest) $M_*$, as one can deduce from panel (a) of
Fig.~\ref{phy_prop}. If we assume the halo mass to scale with the
total baryonic mass ($M_g+M_*$), according to the cosmological ratio,
this implies that more massive galaxies are more efficient in turning
their gas into stars. This trend is as expected, since even a small amount of star formation activity in low mass galaxies can lead to large outflows of gas, thereby suppressing further star formation. This situation however, does not occur in galaxies with larger masses, which do not witness large, galactic scale outflows, by virtue of their much larger potential wells (Mac Low \& Ferrara 1999). 

LAEs with $M_* \leq 10^{9.7} M_\odot$ have $\dot M_* \sim 8-25 \, {\rm
  M_\odot \, yr^{-1}}$, i.e. most of the LAEs have a sustained but not
exceptionally large star formation activity. However, galaxies with
larger stellar masses are much more efficient in converting their gas
into stars, leading to SFR as large as $200 \, {\rm M_\odot\,
  yr^{-1}}$, as seen from panel (b), Fig.~\ref{phy_prop}, which is
just a positive feedback of the $M_g-M_*$ relation mentioned above.

Although the average age of the galaxies are calculated as $t_* = M_* / \dot M_*$, it is interesting to see that all the galaxies we identify as LAEs have $t_*\geq 10$ Myr, as shown in panel (c) of Fig.~\ref{phy_prop}; further the standard deviation on the smallest ages are the smallest. Since we calculate the ages assuming $\dot M_*$ to have been constant over the entire star formation history, it is entirely possible that many of the LAEs are actually older (younger) if $\dot M_*$ in the past was smaller (larger) than the final value at $z \sim 6.1$. However, we are unable to comment on this further in absence of the complete star formation history for each galaxy. Using these average ages, we find that LAEs are intermediate age galaxies, instead of being extremely young ($<$ 10 Myr) or extremely old ($\sim 1$ Gyr) objects. This confirms an earlier result obtained in Dayal et al. (2009a), based on an accurate modelling of the star formation history, that it is unlikely that young ages could be responsible for the large equivalent widths ($EW \geq 200$ \AA) observed for a number of LAEs at $z \sim 4.5, 5.7$ by Dawson et al. (2007) and Shimasaku et al. (2006) respectively. A complete discussion  on the EW distribution is deferred to Sec.~\ref{conc} of this paper.

The mass weighted stellar metallicity of LAEs scales with $M_*$ as shown in panel (d) of Fig.~\ref{phy_prop}; LAEs with a higher $\dot M_*$ are more dust enriched, which is only to be expected since the metals have a stellar origin. The metallicity values for the smallest halos show the largest dispersion, possibly arising due to the differing values of feedback in these low mass halos. Compared to the analogous mass-metallicity relation observed at lower redshifts (Tremonti et al. 2004; Panter et al. 2008; Maiolino et al. 2008), we do not see the sign of a flattening of metallicity towards larger masses which might imply that at high redshift galaxies are not massive enough to retain their metals and behave as close-boxes.

In our model, we have assumed SNII to be the primary dust factories,
with the SNII rate\footnote{The SNII rate is estimated to be $\gamma
  \dot M_* $, where $\gamma \sim (54 \,{\rm M_\odot})^{-1}$ for a
  Salpeter IMF between the lower and upper mass  limits of $1$ and
  ${100\, \rm M_\odot}$ respectively (Ferrara, Pettini and Shchekinov
  2000).} being proportional to $\dot M_*$. In our model, the dust
amount is regulated solely by stellar processes: dust is produced by
SNII, destroyed in the ISM shocked by SNII and astrated into stars, as
mentioned before in Sec.~\ref{dust_model}. The dust amount, therefore,
scales with $\dot M_*$, with the  most star forming galaxies being
most dust enriched, as shown in panel (e) of Fig.~\ref{phy_prop}. A
caveat that must be mentioned here is that as outflows tend to occur
on smaller scales with respect to inflows, they are only marginally
resolved by our RT simulations. Lacking this information, we have not
included their destructive impact into the computation of the dust mass,
which could therefore be somewhat overestimated (refer Dayal et al. 2009b for details). 

As expected, we find that $f_\alpha$ decreases with increasing dust enrichment of the galaxy; galaxies with larger $M_*$, and hence, $\dot M_*$, have a smaller Ly$\alpha$ escape fraction. The value of $f_\alpha$ decreases by a factor of about 3, going from 0.9 to 0.3 as $M_*$ runs from $10^{8.5-10.7} \, {\rm M_\odot}$, as shown in panel (f), Fig.~\ref{phy_prop}.
                                                                                                    
% ***********************************************************************************************
\section{Conclusions and Discussion }
\label{conc}
Although a number of simulations have been used in the past few years (Mc Quinn et al. 2007; Iliev et al. 2008, Nagamine et al. 2008; Dayal et al. 2009a, 2009b; Zheng et al. 2009), the relative importance of reionization, velocity fields and dust, on shaping the observed Ly$\alpha$ and UV LFs has remained rather obscure so far. To this end, we build a coherent model for LAEs, which includes: (a) cosmological SPH simulations run using {\tt GADGET-2}, to obtain the galaxy properties ($\dot M_*$, $Z_*$, $M_*$), (b) the Ly$\alpha$ and continuum luminosities produced, calculated using the intrinsic properties of each galaxy, accounting for both the contribution from stellar sources and cooling of collisionally excited \HI in their ISM, (c) a model for the dust enrichment, influencing the escape fraction of Ly$\alpha$ and continuum photons, and (d) a complete RT calculation, carried out using {\tt CRASH}, including the effects of density and velocity fields, used to calculate the transmission of Ly$\alpha$ photons through the IGM. 

In spite of this wealth of physical effects, we are still left with two free parameters: the radius of dust distribution, $r_d$, which determines the optical depth to continuum photons (see Eq. \ref{tau_dust}) and, the escape fraction of Ly$\alpha$ photons relative to the continuum photons, $f_\alpha/f_c$. These then must be constrained by 
comparing the theoretical model to the observations.

Starting our analysis by assuming all galaxies to be dust-free and ignoring gas peculiar velocities, we find that with such a scheme, none of the ionization states of the IGM with  
$\langle \chi_{HI} \rangle \sim 0.29 $ to $3.4\times 10^{-3}$, can reproduce either the slope or the galaxy number density of the observed UV or Ly$\alpha$ LF. We then include the dust model described in Sec.~\ref{dust_model} into our calculations, such that each SNII produces $0.54 \,{\rm M_\odot}$ of dust, SNII shocks destroy dust with an efficiency of about $12$\% and the dust distribution radius is about half of the gas distribution radius, $r_d =0.48 r_g$. Since the UV is unaffected by the \HI in the IGM as mentioned in Sec.~\ref{model}, the same dust model parameters reproduce the UV LF for all of the ionization states of the IGM, ranging from $\langle \chi_{HI} \rangle \sim 0.29$ to $3.4 \times 10^{-3}$. 

We find that we can reproduce the Ly$\alpha$ LF using $f_\alpha/f_c \sim 1.3$ (3.7) for all ionization states such that $\langle \chi_{HI} \rangle \leq 0.04$ excluding (including) peculiar velocity fields, since the Ly$\alpha$ LF settles to a constant value at this point due to a saturation in $T_\alpha$, as explained in Sec.~\ref{reio effects}. Higher values of $\langle \chi_{HI} \rangle$ naturally lead to a lower transmission, thus leading to a progressive underestimation of the Ly$\alpha$ LF. The higher $f_\alpha/f_c$ value required to reproduce the observations when velocity fields are included, arise because LAEs reside in $\geq 2 \sigma$ fluctuations; 
LAEs are therefore subject to strong inflows which blue-shift the Ly$\alpha$ photons, thereby reducing $T_\alpha$. This decrease in $T_\alpha$ must therefore be compensated by a larger escape fraction from the galaxy to fit the observations. Further, due to the larger masses, $T_\alpha$ is most damped for the largest galaxies, when velocity fields are included, which has the effect of {\it flattening} the slope of the Ly$\alpha$ LF. 

The above {\it degeneracy} between the ionization state of the IGM and
the dust distribution/clumping inside high-redshift galaxies has been
quantified (see Fig.~\ref{prob_contours}); the Ly$\alpha$ LF can be
well reproduced (to within a $5\sigma$ error) by $\langle \chi_{HI}
\rangle \sim0.24$, corresponding to a highly neutral IGM, to a value
as low as $3.4 \times 10^{-3}$, corresponding to an ionized IGM,
provided that the increase in $T_\alpha$ is compensated by a decrease
in the Ly$\alpha$ escape fraction from the galaxy. This leads to the
very interesting conclusion that the ionization state of the IGM can
not be constrained unless the escape fraction of Ly$\alpha$ versus
continuum photons is reasonably well understood. This is possible only
through simulations that model the physics of the ISM, including the
clumping of dust, and include a RT calculation for both Ly$\alpha$ and
continuum photons on such ISM topologies. However, the clumping of dust depends on the turbulence in the ISM, which itself remains only poorly understood. Therefore, attempts to break the mentioned degeneracy must be deferred to future works.

As for the dusty nature of LAEs, we find that $\langle f_c \rangle \sim 0.12$, averaged over all the LAEs in any snapshot, which corresponds to a color excess, $E(B-V) \sim 0.2$, using a SN extinction curve. This value is in very good accordance with the observed values of $E(B-V) < 0.225-0.425$ (Lai et al. 2007), $E(B-V) = 0.035-0.316$ (Finkelstein et al. 2009) and $E(B-V)=0.025-0.3$ (Pirzkal et al. 2007). Secondly, since no single extinction curve (Milky Way, Small Magellanic Cloud, supernova) gives a value of $f_\alpha/f_c >1$; this larger escape fraction of Ly$\alpha$ photons relative to the continuum can be understood as an indication of a multi-phase ISM model (Neufeld 1991) where dust clumps are embedded in a more diffuse ionized ISM component.

LAEs with a larger $M_*$ are also more gas rich. However, the ratio $M_g/M_*$ is the largest (smallest) for LAEs with the smallest (largest) stellar mass, which implies that more massive galaxies ($M_* \geq 10^{10} \, {\rm M_\odot}$) are more efficient in turning their gas into stars ($\dot M_* \geq 40 \, {\rm M_\odot \, yr^{-1}}$ ), i.e., star formation is suppressed ($\sim 8-25 \, {\rm M_\odot \, yr^{-1}} $) in low mass galaxies ($M_* < 10^{9.7} \, {\rm M_\odot}$) due to mechanical feedback. All the galaxies we identify as LAEs have $t_*\sim 10 - 300$ Myr; LAEs are intermediate age galaxies, instead of being extremely young ($<$ 10 Myr) or extremely old ($\sim 1$ Gyr) objects, as shown in Dayal et al. (2009a). The mass weighted stellar metallicity of LAEs scales with $M_*$ but the metallicity of the smallest halos has a large dispersion, possibly arising due to the increasing importance of feedback towards low mass halos. Since in our model, we have assumed SNII to be the primary dust factories, the dust amounts are regulated solely by stellar processes. The dust amounts, therefore, scale with $\dot M_*$, with the  most star forming galaxies being most dust enriched. As expected, we find that $f_\alpha$ decreases with increasing dust enrichment of the galaxy; galaxies with larger $M_*$, and hence, $\dot M_*$, have a smaller Ly$\alpha$ escape fraction. The value of $f_\alpha$ decreases by a factor of about 3, going from 0.9 to 0.3 as $M_*$ increases from $10^{8.5-10.5} \, {\rm M_\odot}$. 

\begin{figure} 
%  \vspace*{10pt} 
  \center{\includegraphics[scale=0.45]{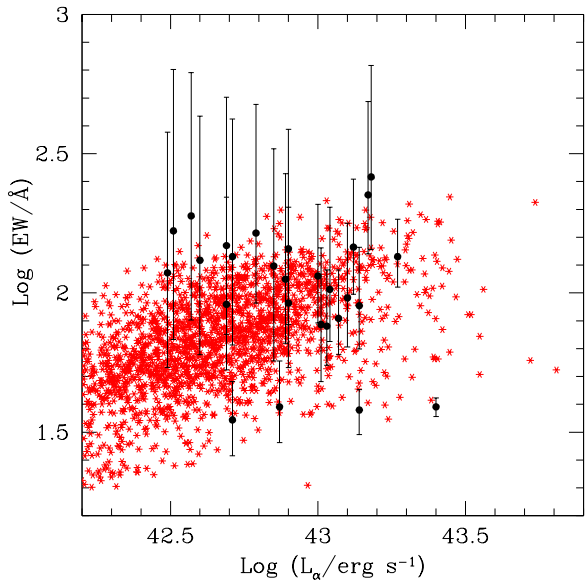}} 
\caption{Observed EWs from Shimasaku et al. (2006) (circles) and model values of the observed EWs (astrexes) as a function of the observed Ly$\alpha$ luminosity. The theoretical model includes the full RT calculation, density, velocity fields, dust and has $\langle \chi_{HI} \rangle \sim 3.4 \times 10^{-3}$, corresponding to model ${\rm M7}$, Tab. \ref{tab1}.}
\label{ew} 
\end{figure}   

We finally comment on the EW distribution obtained from our model and compare it to the observations (Fig.~\ref{ew}). There is a trend of increasing EW in more luminous objects which cannot be yet solidly identified in the available (and uncertain) observational data. In addition, the observed EWs show a large dispersion, varying by as much as a factor of 3 for the same $L_\alpha$ value; this suggests that the observed EW depends on several structural parameters of LAEs, namely $\dot M_*$, peculiar velocities and dust clumping, to name a few. By virtue of using a LAE model which depends on the intrinsic galaxy properties, includes a dust calculation and takes peculiar velocities/ density inhomogeneities into account, we also obtain a large spread (20-222 \AA) in the observed EWs. This is a clear improvement on our own previous work (Dayal et al. 2009) in which some of these ingredients were not yet accounted for.

At the very last, we discuss a few caveats. First, we have used a constant value of the escape fraction of \HI ionizing photons, $f_{esc}=0.02$ in all our calculations. However, the value of $f_{esc}$ remains poorly constrained both observationally and theoretically. While $z \sim 3$ galaxies have been used to obtain values of $f_{esc} < 0.04$ (Fernandez-Soto et al. 2003) to $f_{esc} \sim 0.1$ (Steidel et al. 2001), observations in the local Universe range between $f_{esc} < 0.01$ (Deharveng et al. 1997) to $f_{esc} \sim 0.1-0.73$ (Catellanos et al. 2002). A larger (smaller) value of $f_{esc}$ would lead to larger (smaller) \HII regions, thereby affecting the progress of the reionization process and hence $T_\alpha$; a change in $f_{esc}$ would also affect $L_\alpha^{int}$, Sec.~\ref{intlum_model}.  

Secondly, the average age of the galaxies are calculated as $t_* = M_* / \dot M_*$. Since we calculate the ages assuming constant $\dot M_*$, it is entirely possible that some of the LAEs could be slightly older (younger) if $\dot M_*$ in the past was smaller (larger) than the final value at $z \sim 6.1$. Galaxies younger than $10$ Myr would have higher EWs as compared to the values shown here; however, this could be possible only for the the smallest galaxies, which could assemble due to mergers and undergo star formation in a time as short as 10 Myr.

Third, in the dust model explained in Sec.~\ref{dust_model}, we have considered dust destruction by forward sweeping SNII shocks. However, Bianchi \& Schneider (2007) have shown that reverse shocks from the ISM can also lead to dust destruction, with only about 7\% of the dust mass surviving the reverse shock, for an ISM density of $10^{-25} {\rm gm \, cm^{-3}}$.  Since we neglect this effect, we might be over-predicting the dust enrichment in LAEs. This, however, does not affect our results because the dust optical depth depends on the surface density of the dust distribution as shown in Eq. \ref{tau_dust}; a large (small) dust mass can be distributed in a large (small) volume to obtain identical values of the optical depth.

%*********************************************************************************************
\section*{Acknowledgments} 
We thank C. Evoli, S. Gallerani, R. Schneider and R. Valiante for useful comments. Discussions and the stimulating environment at {\tt DAVID IV}, held at OAArcetri, Florence is kindly acknowledged.

%*********************************************************************************************

\newpage 
\label{lastpage} 
\end{document}